# Description of the shape evolution in the yrast states of $^{186}$Pt

HE Chuang-Ye[1] WU Xiao-Guang[1] ZHENG Yun[1] LI Cong-Bo[1]

1 China Institute of Atomic Energy, Beijing 102413, China

Abstract:

$^{186}$Pt was tested in the framework of IBM-1 and the X(3) model. The results show that $^{186}$Pt is located close to the shape phase transition point, but the B(E2) values little agree with the X(3) model. The shape evolution in the yrast states of $^{186}$Pt is also discussed in detail. TRS calculation exhibits a at bottomed potential at low spin states, but a relatively deep minimum at high spin states. It suggests that a shape evolution from vibrational mode to rotational mode happens in $^{186}$Pt. The result is in agreement with the E-GOS calculation.

Traditionally, nuclear collectivity for the low lying states has been often described in the context of a harmonic vibrator [1], a symmetrically deformed rotor [2], and a deformed γ-soft model [3]. Recently, new models [4, 5], E(5) and X(5), have been proposed for describing nuclei at critical point of shape phase transition between these three ideal structures, in which, the X(5) model [5] describes nuclei at the critical point of the transition from spherical shape to axially symmetric deformed shape. Many even-even nuclei, such as $^{176,178}$Os[6, 7], $^{150}$Nd[8], $^{152}$Sm[9], $^{154}$Gd[10], were tested to have X(5) symmetry. In 2006, a γ-rigid version (with γ= 0) of the X(5), which is called X(3) model, was proposed by Bonatsos Dennis et al [11]. They predicted that $^{172}$Os and $^{186}$Pt may have the X(3) symmetry. But more experimental information needs to be presented. Both of their level energies and B(E2) values should follow the characteristic of X(3) symmetry. One signature of the phase transitional behavior [12, 13] is a sharp rise in the $R_{4/2}$ =E($4^+$)/E($2^+$) value, i.e. the energy ratio between the first $4^+$ state and the first $2^+$ state, as nuclei evolve from the

This work is supported by National Natural Science Foundation of China Grant No. 11175259, 11075214 and 10927507.

vibrator ($R_{4/2}$ = 2.0) to the axial symmetry rotor ($R_{4/2}$ = 3.33). The X(3) solution [11] predicts a value of $R_{4/2}$ = 2.44, which is sitting in between the value for the harmonic vibrator and the deformed rotor.

In Fig. 1 (a), a systematic comparison of R4=2 values is shown for the even even isotopic and isotonic chain of $^{186}$Pt [14, 15]. The isotopic chain shows a smooth evolution in the range of 2.2-2.7 with neutron number increasing except a kink exhibited at $^{176}$Pt, which may result from the local shell effect. It should be noted that $^{182}$Pt has the largest $R_{4/2}$ ratios among the N>98 isotopes. This is probably due to the maximized size of the valence space at the midshell. From the point of view of $R_{4/2}$ ratio, the value of 2.56 for $^{186}$Pt [14] is very close to the value of X(3) model. For the isotonic chain in Fig. 1 (a), the $R_{4/2}$ values display a sudden decreasing, i.e. the nucleus evolves from the deformed rotor to the near harmonic vibrator, as the proton number increases from midshell to the full shell 82. $^{186}$Pt fitly locates at the shape phase transition point in the isotonic chain from deformed rotor to harmonic vibrator. A similar phenomenon is also presented in Fig. 1 (b) for the B(E2) systematic comparisons. With neutron or proton number increasing, both chains of the B(E2) curves show a downtrend for the nuclear mass number larger than 176. The value for $^{186}$Pt [16] just lie in the middle going from $^{180}$Pt to $^{196}$Pt. It gives another evidence that $^{186}$Pt is situated near the critical point of shape phase transition. On the other hand, the potential energy surfaces (PESs) of the $^{184-202}$Pt isotopes [17] exhibit a transition from prolate to oblate shapes between $^{186}$Pt (prolate) and $^{188}$Pt (oblate). The quadrupole moments of $^{184-202}$Pt also exhibit a transition from prolate to oblate behavior between $^{186}$Pt and $^{188}$Pt [17]. Based on the above discussions, $^{186}$Pt is conceived to have the characters of X(3) symmetry.

In Ref [11], the energy spectra of $^{172}$Os and $^{186}$Pt were well reproduced by the X(3) model, especially for $^{186}$Pt. If one nucleus has the symmetry of X(3), its B(E2) values should also follow the characteristic of X(3) symmetry. However, the published B(E2) transition rates for $^{172}$Os [18] have large error bars, and the B(E2) values for $^{186}$Pt were absent for comparison in Ref [11].

Lately, the yrast states of $^{186}$Pt were measured to $16^+$ ℏ by J.C.Walpe et al [16] by using the Doppler-shift recoil distance technique. Their measured values have comparatively low errors relative to $^{172}$Os [18]. Since the energy spectra are predicted to have better agreement with X(3) symmetry than $^{172}$Os, the nucleus $^{186}$Pt is, therefore a good candidate to test the critical point symmetry X(3). The measured results in Ref [16] show a sharp increase in the B(E2) values going from the $2^+$ state to the $6^+$ state. It was interpreted in terms of the mixing of coexisting bands of different deformations at low spins. They have also performed two-band mixing calculations and the calculated results are in good agreement with the observed experimental data. However, the two-band mixing interpretation of these data might not be unique. In this report, we will test the X(3) critical point symmetry in the framework of the interaction boson model (IBM). Furthermore, the shape evolution from low to high spin states along the yrast band of $^{186}$Pt is also discussed in detail.

The interacting boson model, proposed by Arima and Iachello [19], is successful in describing the low lying collective states in many even even nuclei [6,10]. Although the full IBM-1 Hamiltonian for a given nucleus has six parameters, a simplified form reminiscent of an Ising model is almost always used. This Hamiltonian can be written in two fit parameters [20], η and χ. In its basic form, the model describes nuclear excitations and transitions on the basis of bosons, consisting of two coupled valence nucleons, making no distinction between protons and neutrons. In this framework, a standard two dimensional parameterization of the IBM-1 Hamiltonian is

$$\widehat{H} = C[\eta \hat{n}_d - (1-\eta)/N \cdot \widehat{Q}(\chi)\widehat{Q}(\chi)] \qquad (1)$$

where the first term

$$n_d = d^\dagger \cdot \tilde{d} \qquad (2)$$

is the d boson energy and the second is a quadrupole interaction between bosons. The boson quadrupole operator $\widehat{Q}(\chi)$ is given by [21]

$$\hat{Q}(\chi) = (s^\dagger \tilde{d} + d^\dagger s)^{(2)} + \chi \cdot (d^\dagger \tilde{d})^{(2)} \qquad (3)$$

N is the total number of bosons.

Fig. 2 shows the experimental level energies together with the theoretical values of the X(3) model, the symmetric rotor, the vibrator and the IBM-1 fit. The five data sets are scaled by using the experimental $E(2^+)$. Fig. 3 is the same as Fig. 2, but for B(E2).

Though the X(3) model reproduces the experimental spectra very well in Fig. 2, the predicted B(E2: I→I-2)/B(E2:2→0) ratios in Fig. 3 are considerably larger than the experimental values at higher spin states. The experimental ratios are decreasing beyond the $10^+$ state and finally attain the rotor value at 16 ℏ, while the predicted values of X(3) rise monotonously with the spin increasing. It indicates that the nuclei $^{186}$Pt may have the X(3) critical point symmetry at lower spins, but the nuclei evolves into a rotor at higher excited states. Since the X(3) model is a parameter-free prediction utilizing an approximate nuclear potential, it is not surprising that perfect agreement with the experimental data is not obtained. Improved agreement should be possible with a more flexible theoretical model IBM-1. The IBM-1 fitting does improve the agreement for the yrast B(E2) ratios from ground state to $16^+$ ℏ, as shown in Fig. 3. The IBM-1 [20] fitting parameter values are obtained with $\chi = -\sqrt{7/4}$ and $\eta = 0.758$ which are very close to the critical point parameters, but a little closer to the vibrational side. In contrast with the X(3) and IBM-1 models, they both reproduce the experimental energies well. However, in the framework of X(3), it starts to overpredict the experimental B(E2) values from the beginning of 6 ℏ, since the B(E2) ratios of X(3) increase monotonously with the spin; But, the IBM-1 fitting basically agrees with the experimental ratios, although it underestimates the experimental values around 8 ℏ. The difference between X(3) and IBM-1 in this regard stems from a fundamental difference in the philosophy of the two models.

As IBM-1 is originally designed to describe low-spin states of even-even nuclei, it has many difficulties on its applicability in the higher spin regime. In order to better understand the shape evolution in the yrast states of $^{186}$Pt, the empirical ratio of E over spin (E-GOS) curve [22], which is an empirical approach to distinguish vibrational from rotational regimes in atomic nuclei, is also calculated in Fig. 4. For a vibrator, the ideal value of this ratio gradually decreases with spin and inclines to zero, while for an axially symmetric rotor it rises slightly and tends to be a constant value at high spins. For the experimental data of $^{186}$Pt, the ratios of E/I at spins below 14 ℏ are between those for the vibrator and the symmetry rotor, which hints that the nucleus is situated near the critical symmetry point between the spherical and the deformed phase. Though $^{186}$Pt lies near the phase transition point, the E/I ratios gradually diminish to zero as spin increases at low lying levels, which is distinct with a rotor, but very similar to the behavior of vibrators. It suggests that the low spin states are possibly largely built in vibrational mode. However, the E-GOS curve for the states beyond spin 14 ℏ presents a very different tendency. The values tend to be a constant value, then it implies a collective character for the higher spin states. In Fig. 4, the E-GOS ratios from the X(3) model and IBM-1 calculations are displayed for comparison as well. The curve for IBM-1 shows much difference from the experimental data above 10 ~ duo to its limitation in high spin states.

On the other hand, the cranked Woods-Saxon-Strutinsky calculations have been performed as well by means of total Routhian surface (TRS) methods in a three-dimensional deformation space ($\beta_2$, $\beta_4$, $\gamma$) [23,25]. At a given frequency, the deformation of a state is determined by minimizing the resulting total Routhian surfaces. In Fig. 5, (a) and (b) display a prolate shape for the vacuum configuration of $^{186}$Pt nuclei. The minimum locates at $\gamma$ near 0 with a flat bottomed potential. It hints that the deformation is very soft especially in the   direction. Another indicator S(I) defined as [26]

$$S(I, I-1, I-2) = \frac{E(I)+E(I-2)-2E(I-1)}{E(2_1^+)} \quad (4)$$

is a good quantity to deduce if a nucleus is $\gamma$-soft or $\gamma$-rigid deformed by inspecting the relative positions of the even-spin part versus the odd-spin part of a γ band [27]. When the staggering gives positive values for even spins, the potential is triaxial rigid, on the contrary it is γ-soft. The experimental data for the band of $^{186}$Pt can be found in Ref [28]. The calculated S(I) values are presented in Fig. 6. It also shows that $^{186}$Pt is γ softly deformed. It conforms with the TRS calculation. As the bottom of the potential a very shallow shore appears, the shape of nuclei could vibrate at this condition. It gives a reasonable interpretation for the vibrational character at low spin states of $^{186}$Pt displayed in Fig. 4. By comparison with the low spin states, the TRS calculation in Fig. 5 (c) and (d) however, shows a deeper minimum at high spin states. It implies the nucleus has relatively stable deformation. Then the vibrational mode disappears and a rigid rotor behavior may come into being. It is coincident with the E-GOS calculations once again.

An alignment was observed (shown in Fig. 7) at ħω~0.24 MeV [14], in which the Harris parameters used for reference are $J_0$=18 $ħ^2$MeV$^{-1}$ and $J_1$=88 $ħ^4$MeV$^{-3}$. Theoretical quasiparticle energy levels have been calculated for $^{186}$Pt by R. Bengtsson et al [29]. The results show that two quasineutrons from $i_{13/2}$ orbital will be firstly aligned at the frequency of 0.24 MeV. Then the alignment observed in experiment is possibly from those two anti-paired neutrons. It is interesting that the decline of B(E2) values and the inflexion of E-GOS curve happen near the same spin point with alignment. And the results from the TRS calculation in Fig 5 also show that a smaller deformation is induced after band crossing. The shape change in $^{186}$Pt thus could result from the alignment. It was pointed out in Ref [16] that the decline of B(E2) may be caused by a pair of aligned $i_{13/2}$ neutrons. And $^{186}$Pt has 108 neutrons, the quasi-neutrons thus come from the upper part of $i_{13/2}$ orbital. Under such kind of Nilsson orbital filling, the quasi-neutrons induce a stabilization of the shape of $^{186}$Pt, one can see in Fig. 7 the experimental alignments have nearly constant value after band crossing. It apropos supports the results from the E-GOS calculation that $^{186}$Pt has rotational character at high spin states. As the quasiparticles from high j high Ω orbitals have oblate shape driving effect, the two unpaired neutrons from $i_{13/2}$ orbital induce a large γ deformation for $^{186}$Pt at high spin states, as shown in Fig. 5 .

In conclusion, $^{186}$Pt was tested in the framework of the X(3) model. Though $^{186}$Pt is situated very close to the critical point of phase transition fitted by IBM-1, it shows little agreement with the X(3) symmetry. The reason it is possibly due to the large γ softness from the TRS calculation, while the X(3) model is a γ-rigid version of the X(5) model. The E-GOS and TRS calculations have also been performed to study the shape evolution in the yrast band of $^{186}$Pt from low lying states to high spin states, the results suggest that vibrational mode plays a main role at low spin states, but it has a rotational behavior at high spin states. And the shape evolves from a prolate to a large negative deformation.

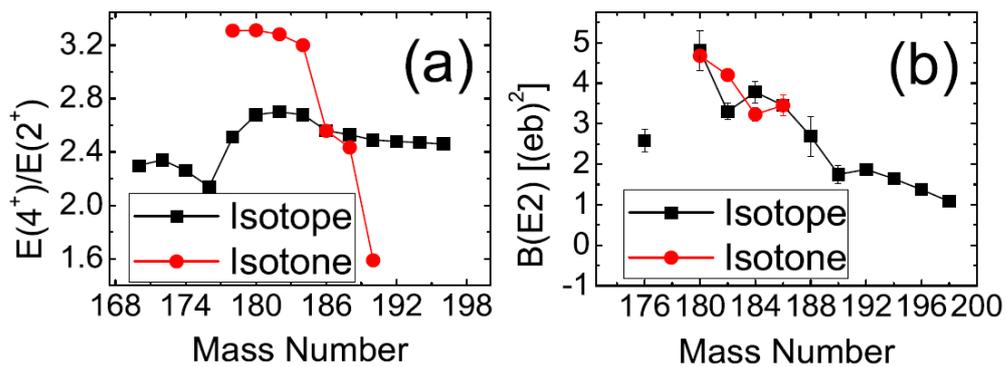

Fig. 1. (Color online) (a) $E(4^+)/E(2^+)$ ratios of the $^{186}$Pt isotopic and isotonic chain. (b) $B(E2:2\rightarrow 0)$ of the $^{186}$Pt isotopic and isotonic chain.

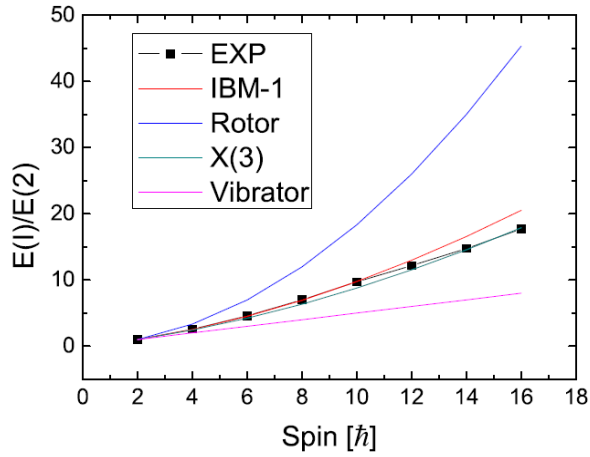

Fig. 2. (Color online)E(I)/E(2) ratios from experimental data of $^{186}$Pt, IBM-1, X(3), ideal rotor and vibrator calculations.

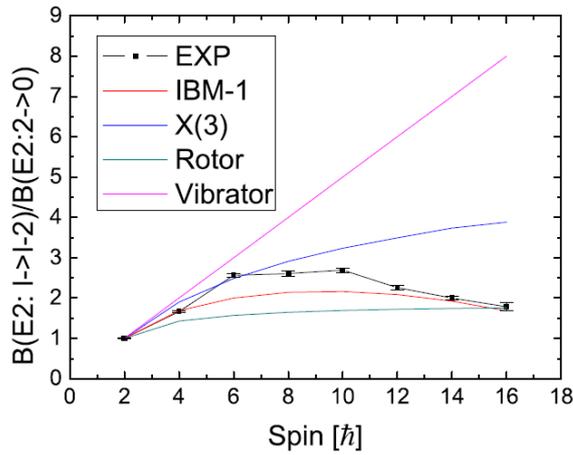

Fig. 3. (Color online)B(E2: I→I-2)/B(E2:2→0) ratios from experimental data of $^{186}$Pt, IBM-1, X(3), ideal rotor and vibrator calculations.

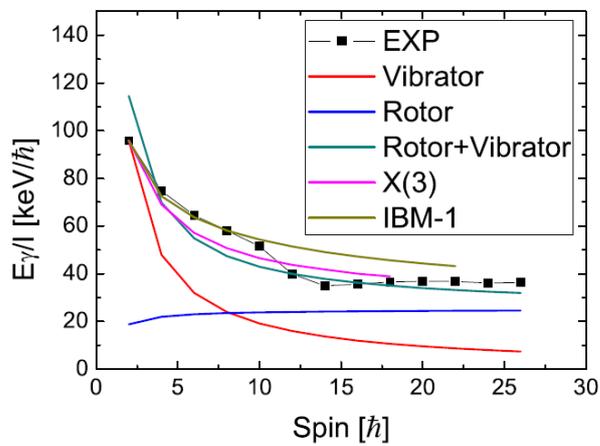

Fig. 4. (Color online)E /I (E-GOS) ratios from experimental data of $^{186}$Pt, IBM-1,

X(3), ideal rotor, ideal vibrator and mixture of rotor and vibrator calculations.

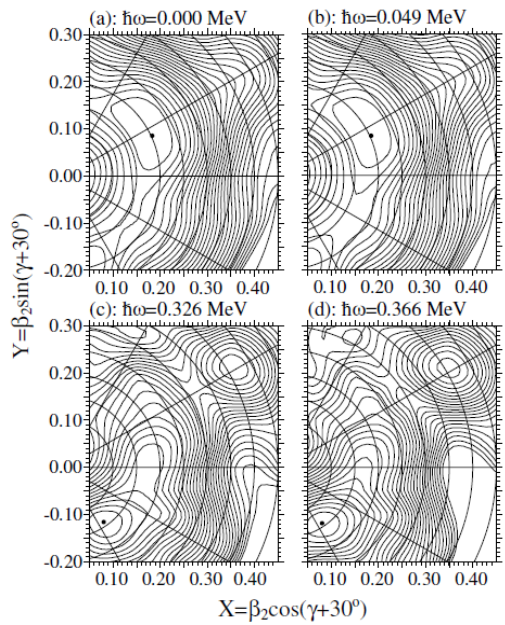

Fig. 5. Polar coordinate plots of total Routhian surface (TRS) for $^{186}$Pt. (a)Vacuum: ℏω = 0.0 MeV, Minimum at β2=0.202, β4=0.040, γ=- 4.9°. (b)Vacuum: ℏω= 0.049 MeV, Minimum at β2=0.203, β4=0.040, γ=-5.4°. (c)Vacuum: ℏω= 0.326 MeV, Minimum atβ2=0.141, β4=0.011, γ=-85.1°. (d)Vacuum: ℏω= 0.366 MeV, Minimum atβ2=0.143, β4=0.008, γ=-85.7°.

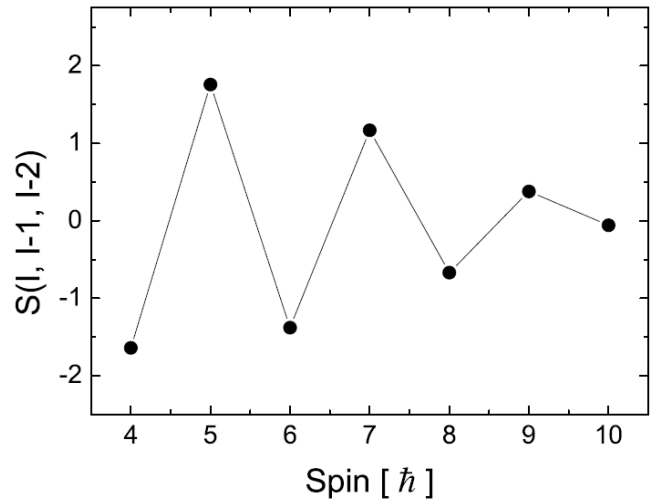

Fig. 6. The experimentally observed odd-even spin energy staggering in the γ band of $^{186}$Pt.

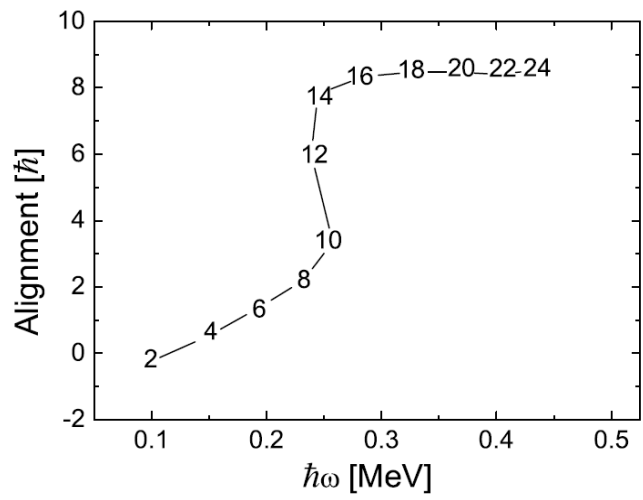

Fig. 7. The experimental alignments of yrast band in $^{186}$Pt. Numbers along the curve indicate spin.